# ARCHITECTURE FOR INTEGRATED MEMS RESONATORS QUALITY FACTOR MEASUREMENT

*Hervé Mathias, Fabien Parrain, Jean-Paul Gilles, Souhil Megherbi, Ming Zhang, Philippe Coste and Antoine Dupret*

Institut d'Electronique Fondamentale UMR 8622 - Université Paris-Sud 11 - Bât. 220
F. 91405 Orsay Cedex – France

**ABSTRACT**

In this paper, an architecture designed for electrical measurement of the quality factor of MEMS resonators is proposed. An estimation of the measurement performance is made using PSPICE simulations taking into account the component's non-idealities. An error on the measured Q value of only several percent is achievable, at a small integration cost, for sufficiently high quality factor values (Q > 100).

## 1. INTRODUCTION

Quality Factor (Q) is one of the most important characteristics of MEMS resonators, especially for vibrating structures where the resonant frequency variation is monitored. For these applications, the higher the Q value, the better the obtained frequency resolution. Quality factor measurement is in these cases very important during testing, since it defines the sensitivity of the corresponding micro-system [1-3].

Within the PATENT DfMM European Network of Excellence, we studied new BIST strategies that would allow the measurement of the MEMS apparent Q factor [4]. As shown on figure 1, the apparent Q factor includes the contributions of the actuation and read-out systems but also the effects due to the packaging or the environment, which play an important role on the effective mechanical quality factor. The aimed Quality Factor Measurement (QFM) techniques could thus be useful at different levels of the component's lifetime: process control monitoring, end-of-line testing; vacuum packaging monitoring, on-line self-test or even auto-calibration, for the aforementioned micro-systems.

In these previous studies, the considered MEMS device has been modeled either as a low-pass 2nd order filter or a band-pass filter, depending on the principle of the movement detection (cf. figure 1). Two promising measurement principles have been identified: Transfer Function Measurement and Step Response Analysis. It has been shown that in terms of accuracy, the second principle is better than the first, which is more suited to the monitoring of Q factor variations [4].

This paper presents the best architecture, among those studied, in terms of accuracy and cost to perform Step Response Analysis. We first present the corresponding measurement principle and the expected performance, then the proposed architecture. Finally, we estimate the effective performance taking into account the components non-idealities.

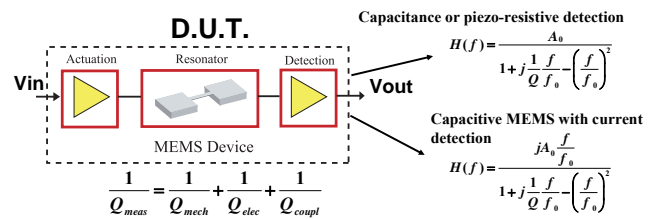

**Figure 1: MEMS devices considered**

$$\frac{1}{Q_{meas}} = \frac{1}{Q_{mech}} + \frac{1}{Q_{elec}} + \frac{1}{Q_{coupl}}$$

Capacitance or piezo-resistive detection:
$$H(f) = \frac{A_0}{1 + j\frac{1}{Q}\frac{f}{f_0} - \left(\frac{f}{f_0}\right)^2}$$

Capacitive MEMS with current detection:
$$H(f) = \frac{jA_0 \frac{f}{f_0}}{1 + j\frac{1}{Q}\frac{f}{f_0} - \left(\frac{f}{f_0}\right)^2}$$

## 2. MEASUREMENT PRINCIPLE

The Step Response Analysis method consists in applying a step actuation on the MEMS structure (like opening the loop in the case of an integrated MEMS-based oscillator) and measuring the amplitude variation of the damped oscillation response at the MEMS output. If the step takes place at t=0s when the oscillator signal is maximal ($V_0$), the obtained damped signal is as follows:

$$V(t) = V_0 \, e^{\left(-\frac{\omega_0}{2Q}t\right)} \left[ \cos\left(\omega_0 \, t \sqrt{1 - \frac{1}{4Q^2}}\right) + \frac{1}{\sqrt{4Q^2 - 1}} \sin\left(\omega_0 \, t \sqrt{1 - \frac{1}{4Q^2}}\right) \right] \quad (1)$$

The measurement's principle consists in counting, in terms of number (n) of elapsed pseudo-periods $T'_0$, the time necessary for the response envelop to move from its initial value $V_0$ down to a fixed voltage $V_0/k$. It is shown on figure 2.

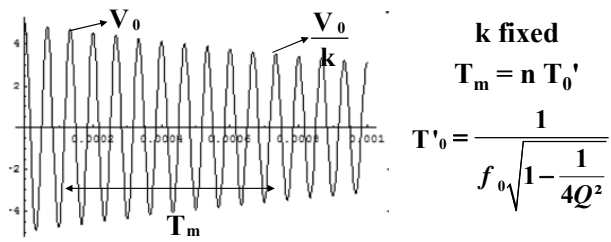

**Figure 2: Fixed Voltage Interval measurement Principle**

From equation (1) we find:

$$k = e^{\left(\frac{\omega_0}{2Q}T_m\right)} \quad (2)$$

From which we obtain the measured value of Q, using the pseudo-period equation given figure 2:

$$Q_{meas} = \frac{\pi f_0}{\ln(k)} T_m = \frac{1}{2}\sqrt{1 + \frac{4\pi^2 n^2}{(\ln(k))^2}} \quad (3)$$

The decaying signal monitoring is performed using a peak detector that measures the values of the signal's successive maxima until the desired value is reached. This method is preferable to the use of an envelope detector. The output of the latter indeed presents ripples that stop the measurement too early, resulting to a decrease in accuracy. The measurement error also depends on the MEMS resonant frequency, which is not the case with the chosen method, at least for a given range of frequencies.

Figure 3 shows the theoretical error obtained with respect to the to-be measured Q value for different values of the division factor k.

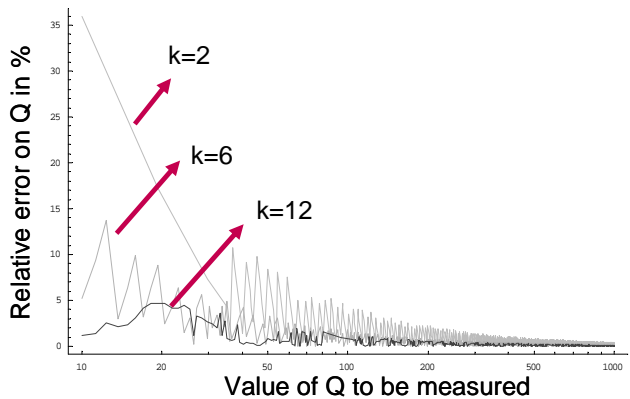

**Figure 3: Theoretical Measurement Error**

The ripples observed on the curves on figure 3 correspond to the fact that the final $V_0/k$ value may not correspond exactly to a maximum of the decaying signal. The number of elapsed pseudo-periods is then an approximation of the effective decaying duration between $V_0$ and $V_0/k$ and the maximum corresponding error on n is 1. For large k values, the number of elapsed pseudo-periods is also large and this error becomes negligible, as can be seen on figure 3. For high Q factors, the theoretical obtained error can be less than 1%.

### 3. PROPOSED ARCHITECTURE

The proposed architecture to perform the integrated Fixed Voltage Interval Measurement is shown on Figure 4. It has been chosen to implement a discrete electronics prototype. The stages used in the final integrated version will probably be different.

Provided its amplitude remains large enough, the decaying signal is used, via a comparator, to generate the clock signal necessary to the control block. The clock signal is also used directly to drive the switch SW4 that resets the peak detector monitoring the decaying signal (C2 capacitor). This way, no *a priori* knowledge of the device's resonant frequency is necessary.

The control block is simple and cheap to implement within an ASIC: it is mainly composed of a counter used to count the desired numbers of pseudo periods. The latter is roughly of the same order as the to-be measured Q values. The control block is also used to drive switches 1 to 3, to launch the step actuation at the beginning of the testing procedure and to compute the effective Q value from the counter output.

The used peak detector requires two wideband opamps allowing the cancellation of the D1 diode threshold while ensuring that the X1 opamp doesn't enter saturation while D1 is off. The components used in simulation have a GBW of 45 MHz and a static voltage gain of 90 dB. The diodes used are small-signal fast diodes. A practical compromise between Gain Bandwidth and input bias current has to be found in the choice of the discrete opamps used: wideband opamps generally have a bipolar input stage that requires a non negligible input bias current that affects the stored voltage on the C2 capacitor. In the ASIC version, the design of a specific wide band OTA should solve this problem. A similar compromise between speed and reverse current has to be found for the diodes.

The voltage divider stage features one low noise, low offset, high input impedance opamp driving a resistance bridge. A comparator is then used to provide the count enable signal to the counter. The counting is stopped as soon as the last detected maximum becomes inferior to the desired threshold.

Figure 5 shows PSPICE simulation results using commercially available components models. The quality factor value is 300 and the k factor is 6, for a resonant frequency of 50 kHz.

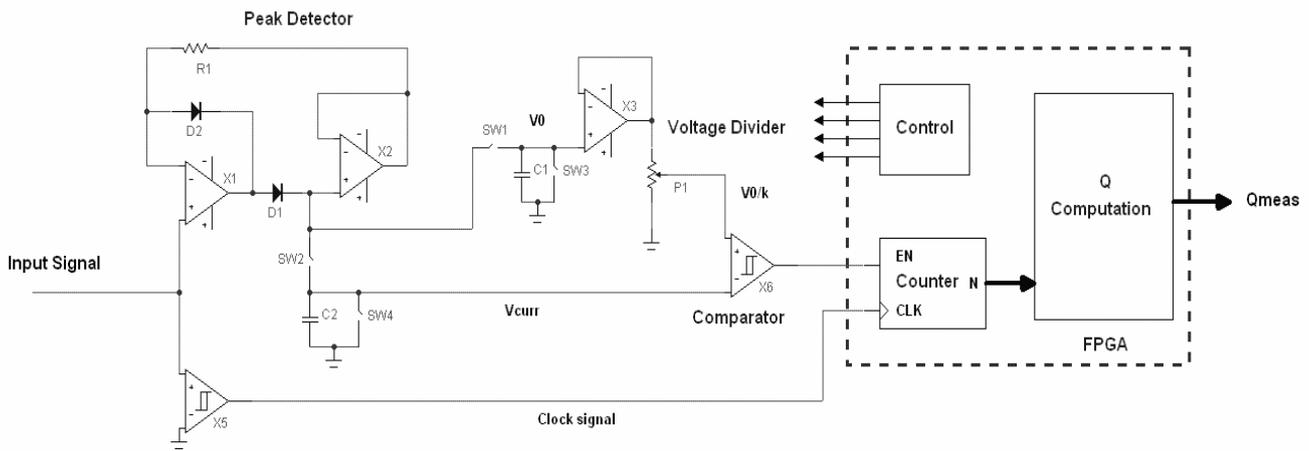

**Figure 4: Prototype architecture for step response measurement**

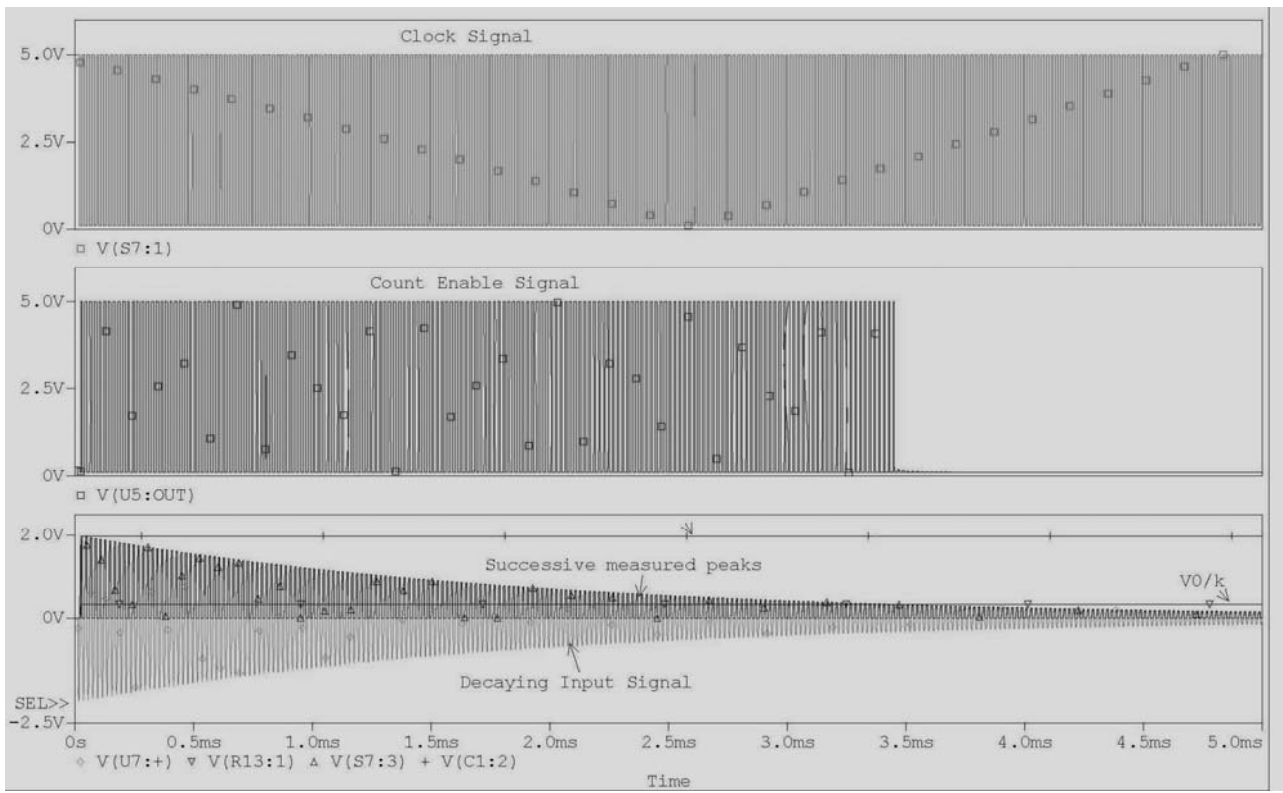

**Figure 5: Prototype architecture PSPICE simulation results**

The measured number of pseudo-periods is 171, corresponding to a measured Q value of 299,8 and thus an error of 0.07%. In this simulation, however, the most important non-idealities (opamps and comparator offsets and noises, technology dispersion impact on k, leakage currents on the capacitors) are not taken into account. Their impact will be estimated and taken into account in the next part.

## 4. PERFORMANCE ESTIMATION

In order to correctly choose the fixed parameters in the architecture (mainly k) and to evaluate the effective performance to be expected, a high level model (Matlab/Simulink) of the architecture has been developed in which the two main sources of measurement error are taken into account: the X5 comparator offset and the error on k.

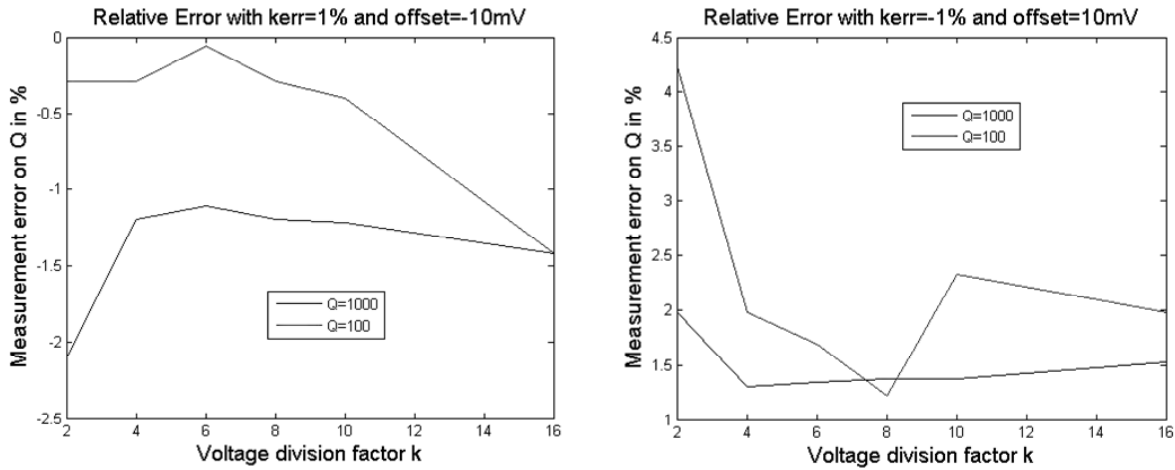

Figure 6: Worst case errors with combined non-idealities

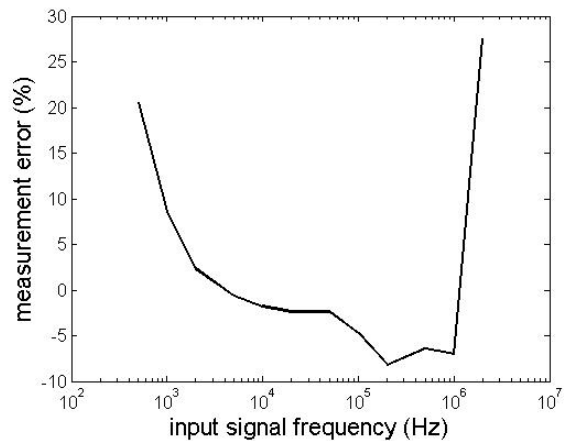

Figure 7: Measurement error with respect to input signal frequency (Q=300)

We chose pessimistic values for these error sources: 1% error on k due to technology dispersions (0.1% is achieved in literature) and 10mV offset for the comparator. Simulations have been performed over a wide range of k and Q values. The worst-case errors are obtained when these two sources induce a threshold variation in the same direction. The corresponding curves are shown Figure 6.

These curves show that, depending on the Q values to be measured, there is an optimal value for k. A choice between 4 and 8 seems a good compromise to be able to accurately measure a wide range of quality factors. For our prototype, we have chosen k=6 to minimize the measurement error. Another value of k would be of interest and will probably be preferred for the ASIC: k=4.81. It is still in the range of k values with small errors and presents the very interesting advantage that the post-processing computation phase is reduced in this case to a simple multiplication by 2 of the counter value. It would thus require no additional hardware.

Further PSPICE simulations have been made taking into account the maximum values of offsets for all the used opamps and comparators and an error on the division ratio of 1% with an input signal with the same quality factor and amplitude as before. The signs of the different offsets have been chosen so as to get the worst-case effects. Different input signal frequencies have been tested to get an insight of the useable frequency range for this architecture. The corresponding measurement errors found are shown on figure 7. They are slightly higher than what was expected with the Matlab/Simulink simulations. It is certainly due to the fact that the offsets of the peak detector and voltage divider opamps cause a further error on the effective k value. A more thourough analysis shows that for low frequencies ($f_0 < 2$ kHz), the main error source is the leakage current at the capacitors. As the frequency increases (2 kHz < $f_0$ < 50 kHz), this error compensates the offsets impact and a very small error can be obtained. Then the offsets impact dominates (50 kHz < $f_0$ < 1 MHz). For higher frequencies (1 MHz < $f_0$), the peak detector fails at canceling the diodes threshold and the error increases greatly. The obtained error remains nonetheless reasonable over a wide range of frequencies: errors of the order of a few percents can be achieved over 3 decades. The results should be improved, both in terms of accuracy and frequency range, with an adapted design of the integrated architecture. Noise has not been taken into account. It should have much less impact than the offsets, being typically 2 orders of magnitude lower.

## 5. CONCLUSION

A very interesting architecture to accurately measure the quality factor of MEMS resonators at low extra cost has been proposed. Its performance taking into account the non-idealities of the components has been estimated and has been shown to be quite good: with pessimistic configurations, the error level is limited to a few percents. A discrete electronics prototype is under development and will soon be used to perform Q factor characterizations. Together with the analysis of error sources performed in this paper, this prototype will give us an important practical feedback in order to properly design the integrated architecture, which will be the next step of this study.

**Acknowledgements:** this work has been funded by the European PATENT – Design for Micro and Nano Manufacture Network of Excellence.